\documentclass[final,twoside,11pt]{entics} 
\usepackage{enticsmacro}

\usepackage[utf8]{inputenc}
\usepackage{enumitem}
\usepackage{proof}
\usepackage{amssymb}
\usepackage{amsmath}
\usepackage{mathrsfs}
\usepackage{mathbbol}
\usepackage{enumitem}
\usepackage{thmtools}
\usepackage{thm-restate}
\usepackage{hyperref}
\usepackage[official]{eurosym}
\DeclareRobustCommand{\officialeuro}{%
  \ifmmode\expandafter\text\fi
  {\fontencoding{U}\fontfamily{eurosym}\selectfont e}}

\usepackage{tikz-cd}

\usepackage{lineno}

\newcommand{\BI}{\textrm{BI}}

\newcommand{\with}{\,\,\&\,\,}

\renewcommand{\emptyset}{\varnothing}
\renewcommand{\phi}{\varphi}

\makeatletter
\def\labelandtag#1#2{\begingroup
   \def\@currentlabel{#2}%
   \phantomsection\label{#1}\endgroup
}
\makeatother

\newcommand{\myhypertarget}[2]{%
  \phantomsection
  \hypertarget{#1}{#2}%
  \expandafter\gdef\csname targettext@#1\endcsname{#2}%
}
\newcommand{\myhyperlink}[1]{%
  \hyperlink{#1}{\csname targettext@#1\endcsname}%
}

\newcommand{\base}[1]{\mathscr{#1}}
    \newcommand{\baseB}{\base{B}}
    \newcommand{\baseC}{\base{C}}

    \newcommand{\baseX}{\base{X}}

\newcommand{\at}[1]{\ensuremath{#1}}
\newcommand{\baseGeq}{\supseteq}

\newcommand{\proves}[1][]{\vdash_{#1}}

\newcommand{\Atoms}{\set{A}}
\newcommand{\Formulas}{\set{F}}
\newcommand{\Bunches}{\set{B}}
    \newcommand{\BunchesWithHole}{\dot{\Bunches}}
\newcommand{\set}[1]{\mathbb{#1}}

\newcommand{\deriveBase}[1]{\vdash_{\!\!#1}}
\newcommand{\deriveBaseM}[1]{\vdash_{\!\!#1}}

\newcommand{\supp}[1]{\Vdash_{#1}}
\newcommand{\suppTor}[1]{\Vdash_{#1}}
\newcommand{\suppM}[2]{\Vdash_{#1 }^{ #2 }}
\newcommand{\suppBI}[2]{\Vdash_{#2 }^{ #1 }}

\newcommand{\At}{\mathbb{A}}
\newcommand{\mand}{\mathrel{\ast}}
\newcommand{\mto}{\wand}
\newcommand{\mtop}{\top^{\ast}}

\newcommand{\addcontext}{\!\mathrel{\fatsemi}\!}
\newcommand{\multcontext}{\!\mathrel{\fatcomma}\!}

    \newcommand{\acomma}{\addcontext}
    \newcommand{\mcomma}{\multcontext}
    \newcommand{\aacomma}{\mathrel{\fatsemi}}
    \newcommand{\mmcomma}{\mathrel{\fatcomma}}

\newcommand{\Multisets}[1]{\mathbb{M}(#1)}

\newcommand{\oc}{!}
\newcommand{\zero}{0}
\newcommand{\one}{1}

\newcommand{\seq}{\triangleright}

\DeclareSymbolFont{bbsymbol}{U}{bbold}{m}{n}
\DeclareMathSymbol{\fatsemi}{\mathbin}{bbsymbol}{"3B}
\DeclareMathSymbol{\fatcomma}{\mathbin}{bbsymbol}{"2C}

    \newcommand{\bunchStrongerThan}{\succeq}

\newcommand{\wand}{\mathrel{-\mkern-12mu-\mkern-12mu\ast}}
\newcommand{\munit}{\varnothing_{\!\times}}
\newcommand{\aunit}{\varnothing_{\!+}}

\newcommand{\rn}[1]{\mathsf{#1}}

\newcommand{\userpwd}{p}
\newcommand{\hwfob}{f}
\newcommand{\OTP}{o}
\newcommand{\access}{s_{\text{acc}}}

\newcommand{\GenRes}{\mathbb{R}}

\makeatletter
\let\orgdescriptionlabel\descriptionlabel
\renewcommand*{\descriptionlabel}[1]{%
  \let\orglabel\label
  \let\label\@gobble
  \phantomsection
  \edef\@currentlabel{#1}%
  \let\label\orglabel
  \orgdescriptionlabel{#1}%
}
\makeatother

\newcommand{\quotes}[1]{\text{`$#1$'}}

\usepackage{graphicx}

\def\conf{MFPS 2024}
\def\myconf{mfps40} 	
\volume{4}			
\def\volu{4}			
\def\papno{9}			
\def\mydoi{14727}%
\def\lastname{Gheorghiu, Gu, Pym} 
\def\runtitle{Inferentialist Resource Semantics} 
\def\copynames{A. V. Gheorghiu, T. Gu, David. J. Pym}    

\begin{document}

\begin{frontmatter}

\title{Inferentialist Resource Semantics \\
        (Extended Abstract)\thanksref{funding}\thanksref{people}}

 \thanks[funding]{This work has been partially supported by the UK EPSRC grants EP/S013008/1 and EP/R006865/1 and by the EU MOSAIC MCSA-RISE project.}
 \thanks[people]{We are grateful to Gabriele Brancati, Tim Button, Yll Buzoku, Tristan Caulfield, Diana Costa, Timo Eckhardt, Didier Galmiche, Timo Lang, Sonia Marin, Peter O'Hearn, Elaine Pimentel, Eike Ritter, Edmund Robinson, and Will Venters for discussions of various aspects of this work.}
        
 \author{Alexander V. Gheorghiu\thanksref{a}\thanksref{alexemail}}	
   \author{Tao Gu\thanksref{a}\thanksref{taoemail}}
   \author{David J. Pym\thanksref{a}\thanksref{b}\thanksref{c}\thanksref{davidemail}}	
   \address[a]{Department of Computer Science \\ University College London \\ London WC1E 6BT, UK}  				
   
   \thanks[alexemail]{Email: \href{mailto:alexander.gheorghiu.19@ucl.ac.uk} {\texttt{\normalshape
        \textbf{alexander.gheorghiu.19@ucl.ac.uk}}}} 

         \thanks[taoemail]{Email: \href{mailto:tao.gu.18@ucl.ac.uk} {\texttt{\normalshape
        \textbf{tao.gu.18@ucl.ac.uk}}}}

      \address[b]{Department of Philosophy \\ University College London \\ London WC1E 6BT, UK}  				

  \address[c]{Institute of Philosophy \\ University of London \\ London WC1H 0AR, UK} 
  \thanks[davidemail]{Email:  \href{mailto:d.pym@ucl.ac.uk} {\texttt{\normalshape
        d.pym@ucl.ac.uk}}}

\begin{abstract} 
In systems modelling, a \emph{system} typically comprises located resources relative to which processes execute. One important use of logic in informatics is in modelling such systems for the purpose of reasoning (perhaps automated) about their behaviour and properties. To this end, one requires an interpretation of logical formulae in terms of the resources and states of the system; such an interpretation is called a \emph{resource semantics} of the logic. This paper shows how \emph{inferentialism} --- the view that meaning is given in terms of inferential behaviour ---  enables a versatile and expressive framework for resource semantics. Specifically, how inferentialism seamlessly incorporates the assertion-based approach of the logic of Bunched Implications, foundational in program verification (e.g., as the basis of Separation Logic), and the renowned number-of-uses reading of Linear Logic. This integration enables reasoning about shared and separated resources in intuitive and familiar ways, as well as about the composition and interfacing of system components. 
\end{abstract} 

\begin{keyword}
  inferentialism, proof-theoretic semantics, resource semantics, linear logic, bunched logic, systems modelling
\end{keyword}

\end{frontmatter}

\section{Introduction}\label{sec:intro}

 Within informatics, perhaps the most important systems concept is that of a \emph{distributed system}~\cite{CDKB-DS-2011,Tannenbaum2001}. From the systems modelling perspective, with a little abstraction, one can think of a distributed system as comprising delimited collections of interconnected component systems~\cite{Milner2002,Tripakis2011,CP15,CMP2012,AP16,GLP2024}. These systems ultimately comprise \emph{locations}, at which are situated \emph{resources} 
 relative to which \emph{processes} execute, consuming, creating, moving, combining, and otherwise manipulating resources as they evolve, so delivering services.


One primary application of logic in informatics is for representing, understanding, and reasoning about systems; this determines the field of \emph{logical systems modelling}. In this context, the `modelling' is used both in the general and mathematical logical sense. The goal is to utilize logic to represent, analyze, and simulate systems by interpreting logical structures and relationships in terms of concepts relevant to the model in question. We discuss several examples below. 

In the field of logical systems modelling, \emph{substructural logics} are useful because of their \emph{resource interpretations}. The study of such interpretations of logics, especially in the context of systems modelling, is called \emph{resource semantics} --- see Section~\ref{sec:res-int}. 
Perhaps the most celebrated examples of resource semantics are the `number-of uses' reading of \emph{Linear Logic} (LL)~\cite{girard1987linear} and the `sharing/separation' reading of \emph{the logic of Bunched Implications} (BI)~\cite{o1999logic}. While both are applicable to systems modelling, these two readings work in different ways and are related to different ways of doing the modelling as we discuss presently. 

The \emph{number-of-uses reading} of LL  (e.g.,~\cite{Lafont,Girard,Abramsky,Hoare}) proceeds through LL's proof theory and is not reflected in its truth-functional semantics (e.g., \cite{Girard,Allwein,COUMANS201450}). Heuristically, it concerns the dynamics of the system: formulae denote processes and resources themselves. It is conveniently 
illustrated by the (by now familiar) \emph{Vending Machine Model}:

\begin{quote}
\begin{quote}
\begin{quote}
\emph{Having a chocolate bar is denoted by 
$C$, and having one euro by $1{  \euro}$.  The formula $1{ \euro} \to C$ (using material implication) is intended to denote `$1{ \euro}$ buys one chocolate bar', since (by \emph{modus ponens}) $1\euro$ together with $1{  \euro} \to C$ yield $C$. However this probably doesn't model the economy we meant as it also follows that $1{  \euro} \to C\land C$ (i.e., `$1{ \euro}$ buys two chocolate bars')! The point is that a euro is a resource that should be \emph{consumed} in a transaction, yielding only one chocolate bar. We can model things more carefully using linear implication and the tensor product: from $1{  \euro} \multimap C$, we infer $1{  \euro} \otimes 1{  \euro}   \multimap C \otimes C$, but not $1{  \euro} \multimap C \otimes C$.}  
\end{quote}
\end{quote}
\end{quote}
Observe that we have not given a model in the formal sense of an algebraic structure, but rather have given a reading of the logical structures and relationships within LL in terms of some concepts pertinent to vending machines. 

In contrast to the number-of-uses reading of LL, the \emph{sharing/separation} reading of BI (e.g.,~\cite{o1999logic,pym2019resource,AP16,pym2019separation})  proceeds through its (order\-ed-monoidal/relational) model-theoretic semantics \cite{GP2023-SABI}. In this case, one has models in the general sense represented by models in the algebraic sense; for example, beginning with the stack-heap model of computer memory and using the \emph{sharing interpretation} of BI, Ishtiaq and O'Hearn~\cite{Ishtiaq2001} (following Reynolds~\cite{reynolds2000intuitionistic}) invented the \emph{stack-heap} model of (usually Boolean) BI on which Separation Logic~\cite{reynolds2002separation} rests. The details of this and its efficacy has been 
discussed by Pym et al.~\cite{pym2019resource,pym2019separation,Galmiche2019ASE}. Returning to the \emph{Vending Machine Model} above:



\begin{quote}
    \begin{quote}
        \begin{quote}
  \emph{We use the ordered monoid of the natural numbers $\langle \mathbb{N}, \leq, +, 0\rangle$ to model euros. Suppose chocolate bar $A$ costs $2 \euro$ and chocolate bar $B$ costs $3 \euro$: we write $3 \models A \land B$ to say that $3 \euro$ suffice for \emph{both} chocolates in the sense that one could purchase either one; and we write $5 \models A  \mand B$ to denote that $5 \euro$ may be split/shared to purchase $A$ and $B$ \emph{separately}. Note that, in the first example ($\land$), we use \emph{persistence}, in the sense that as $2 \euro$ suffice for $A$ and $3>2$, so do $3 \euro$.} 
        \end{quote}
    \end{quote}
\end{quote}

There are some important facts about the number-of-uses and sharing/separation readings and their use in systems modelling that need to be emphasised: 
\begin{itemize}[nosep,label=--]
    \item firstly, while both readings are useful, they are individually limited in the context of systems modelling: sharing/separation expresses the \emph{structure} of distributed systems and number-of-uses expresses the \emph{dynamics} of the resources involved;  
    \item secondly, the two readings operate in completely different paradigms: number-of-uses proceeds from the proof theory of LL, while sharing/separation proceeds from the model theory of BI. 
    \end{itemize}
    We propose and discuss the use of a unifying framework for the two resource interpretations given above: \emph{proof-theoretic semantics} (P-tS). The semantic paradigm supporting P-tS is \emph{inferentialism}~\cite{Brandom2000} --- the view that meaning (or validity) arises from inference. Therefore, as a point of departure for this work, we adopt an  inferentialist view of systems modelling in which the basis for reasoning about a system and its properties should be based on inferences over the underlying specification of the system. 
    More precisely, in distributed systems \emph{policies} are used to determine behaviours. From the inferentialist point of view, these behaviours can be interpreted as (collectively) giving the \emph{meaning} of the distributed system \cite{jaakko}. In this paper, we give an account of such `inferential' models of distributed systems using recent advances in the proof-theoretic semantics of substructural logics. For example, an implication $\phi \to \psi$ denotes a policy for an action that moves the system from a state that satisfies policy $\phi$ to a state that satisfies policy $\psi$: instantiated to the Vending Machine Model, policy $\phi$ is the state of having $1 \euro$ and policy $\psi$ is the state of having a chocolate bar, our model of the policy should make the exchange of resource linear in the sense that, having executed it, one no longer has the resource $1\euro$. Details are given in Section~\ref{sec:res-int}.

Having given the relevant background in proof-theoretic semantics in Section~\ref{sec:b-es}, the main contribution of the paper begins in Section~\ref{sec:res-int} with a general definition of \emph{resource semantics}. We explain how the recent P-tS of LL and of BI, respectively, correspond to their number-of-uses and sharing/separation interpretations; 
briefly, they both proceed by a semantic judgment relation called \emph{support}, $\Gamma \suppM{\baseB}{R} \phi$. In Section~\ref{sec:res-int}, we give an interpretation of each component of support judgments in the context of systems modelling: intuitively, $\Gamma$ 
describes a system policy, $\baseB$ represents an inferential model of a system policy, $R$ constitutes a collection of available resources, and $\phi$ signifies an assertion regarding the executed system according to the policy of $\Gamma$ and $\baseB$ and collection resources $R$. In Section~\ref{sec:example}, we offer an illustration of this interpretation by modelling the departure security infrastructure of an airport and multi-factor authentication. These examples are intended to be both familiar and evidently generic in that it should be clear how other examples would map onto them quite naturally.
In Section~\ref{sec:generality}, we discuss informally how our inferentialist resource semantics applies to distributed systems in general. 

We include a brief glossary for the various logics discussed in this paper. We have several subsystems of \emph{Linear Logic} (LL): \emph{intuitionistic} LL is abbreviated ILL, ILL without $\oc$ is abbreviated IMALL, IMALL without $\with, \oplus, 1, \bot$ is abbreviated IMLL. The \emph{logic of Bunched Implications} is abbreviated BI. Finally, \emph{Intuitionistic Propositional Logic} is abbreviated IPL.



\section{Proof-theoretic Semantics: Base-extension Semantics} \label{sec:b-es} 

In model-theoretic semantics (M-tS), logical consequence is defined in terms of truth in models --- understood as abstract mathematical structures. In the standard reading of Tarski~\cite{Tarski1936,Tarski2002}, a propositional formula $\phi$ follows (model-theoretically) from a context $\Gamma$ iff every model of $\Gamma$ is a model of $\phi$.
Therefore, consequence is understood as the \emph{transmission of truth}. 
From this perspective, \emph{meaning} and \emph{validity} are characterized in terms of \emph{truth}. 

Meanwhile, in proof-theoretic semantics (P-tS)~\cite{Prawitz1971ideas,prawitz1973towards,Dummett1991logical,Francez2015,wansing2000idea,Alex:PtV-BeS}, meaning and validity are characterized in terms of \emph{proofs} --- understood as objects denoting collections of acceptable inferences from accepted premisses --- and \emph{provability}. To emphasize: it is not that one provides a proof system for the logic, but rather one explicates the meaning of the connectives \emph{in terms of} proof systems. Indeed, as Schroeder-Heister \cite{Schroeder2007modelvsproof} observes, since no formal system is fixed (only notions of inference) the relationship between semantics and provability remains the same as it has always been: soundness and completeness are desirable features of formal systems. Essentially, \emph{proof} in P-tS plays the part of \emph{truth} in M-tS.  

The field of P-tS  is wide and encompasses several distinct approaches. In this paper, we restrict attention to \emph{base-extension semantics} (B-eS), which is a particular approach to P-tS. A B-eS is a characterization of a logic by a judgment relation called \emph{support} that defines the validity of formulae. It is inductively defined according to the syntax of the logic. It is analogous to the \emph{satisfaction} judgment in M-tS (cf. \cite{kripke1965semantical}). Crucially, the base case is given by `derivability in a base', which is regarded as a \emph{logic-free} notion of proof; that is, bases are proof systems restricted to atoms. Despite being structurally similar to M-tS, the subtle differences in the set-up have significant consequences (discussed below).
 
In this paper, we follow the approach to B-eS by Sandqvist~\cite{Sandqvist2005inferentialist,Sandqvist2009CL,Sandqvist2015base} and Piecha et al.~\cite{Piecha2015failure,Piecha2016completeness,Piecha2019incompleteness}. While there are other versions than that presented here (cf.~\cite{goldfarb2016dummett,Stafford2023}), they are not directly relevant for this work. 
 In this section, we present three B-eS on which we base the inferentialist account of resource semantics. First, in Section~\ref{sec:b-es:ipl}, we discuss the B-eS of \emph{intuitionistic propositional logic} (IPL) as this provides a good background and example of how B-eS works. Second, in Section~\ref{sec:b-es:ll}, we discuss the B-eS of (inutitionistic) LL, which arises through modifications on the B-eS of LL by accounting for some notion of atomic `resource'. Finally, in Section~\ref{sec:b-es:bi}, we discuss the B-eS of BI, which enriches the work on LL with the more delicate  structure of bunches. 

\subsection{Intuitionistic Propositional Logic} \label{sec:b-es:ipl}

In this section, we introduce the techniques of B-eS through a treatment, due to Sandqvist \cite{Sandqvist2015base}, of IPL. Fix a (denumerable) set of atomic propositions $\At$. Here (and in the sequel) lower-case Romans denote atoms and upper-case Romans denote sets (and, later, multisets and bunches) of atoms; lower-case Greeks denote formulae and upper-case Greeks denote sets (and, later, multisets and bunches) of formulae. 

The B-eS for IPL begins by defining atomic rules. An \emph{atomic rule} is a natural deduction rule of the following form, in which $p,p_1,...,p_n$ are atoms and $P_1$,...,$P_n$ are (possibly empty) sets of atoms: 
\[{
 \infer[\rn{A}]{~~\at{p}~~}{}\qquad 
    \infer[\rn{R}]{\at{p}}{~~\deduce{\at{p}_{1}}{[P_{1}]} & ... & \deduce{\at{p}_n}{[P_n]}~~} } 
\]

A \emph{base} is a set of such atomic rules. We write $\baseB, \baseC, \dots$ to denote bases, and $\emptyset$ to denote the empty base (i.e., the base with no rules). We say $\base{C}$ is an \emph{extension} of $\base{B}$ if $\baseC$ is a superset of $\baseB$, denoted $\base{C} \supseteq \base{B}$. 

Importantly, atomic rules are taken \emph{per se} and not closed under substitution when creating derivations.

\begin{defn}[Derivability in a Base]  \label{def:derivability-base-IPL}
\emph{Derivability in a base} $\baseB$ is the smallest relation $\proves[\baseB]$ satisfying the following:

\begin{itemize}[nosep,label=--]
    \item \textsc{ref:} $P, p \deriveBase{\baseB} p$, for any $p \in \At$ 
    \item \textsc{app$_1$:} If $\rn{A} \in \baseB$, then $P \deriveBaseM{\baseB} p$
    \item \textsc{app$_2$:} If $\rn{R} \in \baseB$ and $P, P_1\deriveBase{\baseB} p_1$,\ldots, $P,P_n\deriveBase{\baseB} p_n$, then $P \deriveBaseM{\baseB} p$.
\end{itemize}
\end{defn}

Derivability in a base ($\deriveBase{\baseB}$) is the base case of \emph{support in a base} ($\supp{\baseB}$) that gives the B-eS --- that is, for any $p \in \At$, $\supp{\baseB} p$ iff $\deriveBase{\baseB} p$. The inductive case is given analogously to the definition of satisfaction in M-tS. 

\begin{defn}[Support for IPL] \label{def:BeS-Tor}
Support is the smallest relation $\suppTor{}$ satisfying the clauses of Figure~\ref{fig:sandqvist}.
\end{defn}

\begin{figure}[t]
\hrule \vspace{1mm}
 \[ 
        \begin{array}{l@{\qquad}c@{\quad}l@{\quad}r}
            \suppTor{\base{B}} \at{p}  & \mbox{iff} &   \proves[\base{B}] \at{p} & \mbox{(At)}  \\[-.5mm]
             \suppTor{\base{B}} \phi \to \psi & \mbox{iff} & \phi \suppTor{\base{B}} \psi & (\to) \\[-.5mm]
             \suppTor{\base{B}} \phi \land \psi   & \mbox{iff} & \mbox{$\suppTor{\base{B}} \phi$ and $\suppTor{\base{B}} \psi$} & (\land) \\[-.5mm]
             \suppTor{\base{B}} \phi \lor \psi & \mbox{iff} &  \mbox{$\forall \base{C} \supseteq \baseB\, \forall \at{p} \in \At$, if $\phi \suppTor{\base{C}} \at{p}$ and  $\psi \suppTor{\base{C}} \at{p}$, then $\suppTor{\base{C}} \at{p}$} & (\lor)  \\[-.5mm]
             \suppTor{\base{B}} \bot & \mbox{iff} &    \suppTor{\base{B}} \at{p} \text{ for any } \at{p} \in \At & (\bot) \\[-.5mm]
            \suppTor{\baseB} \Gamma & \mbox{iff} & \suppTor{\baseB} \gamma \text{ for any } \gamma \in \Gamma \\[-.5mm]
           \hspace{-1em} \Gamma \suppTor{\base{B}} \phi & \mbox{iff} & \mbox{$\forall \base{C} \supseteq \baseB$, if $\suppTor{\base{C}} \Gamma$, then $\suppTor{\base{C}} \phi$ } &  \mbox{(Inf)} \\[-.5mm]
            \hspace{-1em} \Gamma \suppTor{} \phi & \mbox{iff} & \mbox{$\Gamma \suppTor{\baseB} \phi$ for all bases $\baseB$} &  
        \end{array}
        \]
    \hrule
    \caption{Base-extension Semantics for IPL} 
    \label{fig:sandqvist}
\end{figure}

\begin{theorem}[Sandqvist~\cite{Sandqvist2015hypothesis}]
    $\Gamma \proves[\emph{IPL}] \phi$ iff $\Gamma \supp{}{} \phi$.
\end{theorem} 

Soundness --- that is, $\Gamma \proves[\textrm{IPL}] \phi$ implies $\Gamma \supp{}{} \phi$ --- follows from showing that support admits all the rules of Gentzen's \textsc{NJ} \cite{Gentzen}. Completeness --- that is, $\Gamma \supp{}{} \phi$ implies  $\Gamma \proves[\textrm{IPL}] \phi$ --- is more subtle. In this case one constructs a special base $\base{N}$ that `simulates' Gentzen's $\textsc{NJ}$~\cite{Gentzen} and uses the clauses of the semantics to show that support in $\base{N}$ is indeed equivalent to provability in $\textsc{NJ}$.

Before concluding this section, there are a couple of important features of the semantics to note:

\begin{remark}
The clauses for various connectives in a B-eS can be substantially different from their treatments in other semantics. In particular, the B-eS for IPL has the following clauses for disjunction ($\lor$):
\[ {
 \suppTor{\base{B}} \phi \lor \psi \qquad \mbox{iff} \qquad \mbox{$\forall \base{C} \supseteq \baseB\, \forall \at{p} \in \At$, if $\phi \suppTor{\base{C}} \at{p}$ and  $\psi \suppTor{\base{C}} \at{p}$, then $\suppTor{\base{C}} \at{p}$} }
\]

\noindent This is one significant consequence of the inferentialist set-up: Piecha et al.~\cite{Piecha2015failure,Piecha2016completeness,Piecha2019incompleteness} have shown that the standard meta-level `or' clause  --- that is, $\supp{\baseB} \phi \lor \psi$ iff $\supp{\baseB} \phi$ or  $\supp{\baseB} \psi$ --- yields incompleteness. Note that the clause is closely related to the `second-order' definition of disjunction (see Prawitz~\cite{Prawitz2006natural}) --- that is, 
\[
    {
    U+V = \forall X \left((U \to X) \to (V \to X)\to X\right)
    }
\]
\normalsize
 This adumbrates the categorical perspective on B-eS for IPL given by Pym et al.~\cite{Pym2022catpts,PRR2024}
\end{remark}
\begin{remark}
The treatment of non-empty context is given by a \emph{base-extension},
\[
    {
    \mbox{$\Gamma \supp{\baseB} \phi$} \qquad \mbox{iff} \qquad \mbox{$\forall \baseX \supseteq \baseB$, if  $\supp{\baseX} \Gamma$, then $\supp{\baseX} \phi$}
    }
\]
While \emph{prima facie} this use of base-extension appears to be related to pre-order in the M-tS of IPL (cf.~\cite{kripke1965semantical}), that remains to be determined (though see \cite{goldfarb2016dummett,Stafford2023}).
\end{remark}

This brief introduction suffices to highlight some important features of B-eS. The remainder will now give the detail in the context of substructural logic.

\subsection{Linear Logic} \label{sec:b-es:ll}

The challenges and motivations involved in producing a B-eS for a substructural logic are discussed in detail by the authors~\cite{GGP2023-IMLL,GGP2024-BI}. In this section, we present the treatment of LL provided by the authors~\cite{GGP2023-IMLL} (with the additives given by Buzoku~\cite{Yll-arxiv,Yll-bristol}).  

It is instructive to begin with the semantics for IMLL. This simplifies the problem to the following: \emph{How does one make support for IPL linear?} Looking at (Inf) in Figure~\ref{fig:sandqvist}, we expect that $\phi$ being supported in a base $\baseB$ relative to some multiset of formulas $\Gamma$ means that the `resources' garnered by $\Gamma$ suffice to produce $\phi$. We express this by enriching the notion of support with multisets of resources $P$ and $U$ combined with multiset union --- denoted $\,\mcomma\,\,\,$ --- as follows:
\[ {
\begin{array}{l@{\quad}c@{\quad}l}
    \Gamma \suppM{\baseB}{S} \phi & \mbox{iff} & 
    \mbox{for any $\baseX \baseGeq \baseB$ and any $\at{U}$, if $\suppM{\baseX}{\at{U}} \Gamma$, then $\suppM{\baseB}{\at{S} \mmcomma \, \at{U}} \phi$}
     \end{array} }
\]
where
\[ {
\begin{array}{l@{\quad}c@{\quad}l}
    \suppM{\baseB}{U} \Gamma_1 \mcomma \Gamma_2 & \mbox{iff} & \mbox{there are $U_1$ and $U_2$ such that $U = (U_1 \mcomma U_2)$, $\suppM{\baseX}{\at{U}_1} \Gamma_1$, and  $\suppM{\baseX}{\at{U}_2} \Gamma_2$}
\end{array} }
\]
The `resource' reading here is indeed inspired by the number-of-uses reading of LL. In this way, this paper presents a semantics of ILL in which this celebrated reading is, by design, present. 

We now continue to give the technical details of semantics.  Again, fix a (denumerable) set of atoms $\At$. Recall that we use lower-case Roman letters $p,\ldots$ to denote elements of $\At$, upper-case Roman letters $P,Q,S,U,V\ldots$ to denote (finite) \emph{multisets} of elements of $\At$, lower-case Greek letters $\phi, \psi, \chi,\ldots$ to denote formulae, and upper-case Greek letters $\Gamma, \Delta,\ldots$ to denote (finite) \emph{multisets} of formulae. More generally, for a set $X$, we write $\Multisets{X}$ to denote the set of (finite) \emph{multisets} of elements of $X$; we use $\mmcomma$ to denote multiset union, and $\emptyset$ to denote the empty multiset.

For ILL, atomic rules are taken in the format used by Bierman~\cite{Bierman1994intuitionistic} and Negri~\cite{Negri2002}; that is, rules have premisses with (possibly empty) multiset of hypothesises that are divided into sets,
\[ {
\infer[\rn{A}]{~~p~~}{} \qquad 
\infer[\rn{R}]{p}{\left\{\raisebox{-1em}{$\deduce{p_1^{(1)} \quad \ldots \quad p_{n_1}^{(1)}}{\,\,[P_1^{(1)}] \hspace{2.8em} [P_{n_1}^{(1)}]}$}\right\} &\ldots & \left\{\raisebox{-1em}{$\deduce{p_1^{(k)} \quad \ldots \quad p_{n_k}^{(k)}}{\,\,[P_1^{(k)}] \hspace{2.8em} [P_{n_k}^{(k)}]}$}\right\}} }
\]
A base $\baseB$ is a set of such rules; we use the same conventions as in Section~\ref{sec:b-es:ipl}.

\begin{defn}
    [Derivability in a Base] ~\label{def:derivability-base-LL} Derivability in a base $\baseB$ is the smallest relation $\proves[\baseB]$ satisfying the following:
\begin{itemize}[nosep,label=--]
    \item \textsc{ref:} $p \deriveBase{\baseB} p$, for all  $p \in \At$
    \item \textsc{app$_1$:} If $\rn{A} \in \baseB$, then $\emptyset \deriveBase{\baseB} p$.
    \item \textsc{app$_2$:} If $\rn{R} \in \baseB$ and there are $C_1,\ldots  ,C_k \in \Multisets{\At}$ s.t. $C_i \mcomma P_j^{(i)} \deriveBase{\baseB} p_j^{(i)}$ for every $i=1 \ldots k$ and $j=1 \ldots n_i$, then $C_1 \mcomma \ldots \mcomma C_k \deriveBase{\baseB} p$.
\end{itemize}
\end{defn}
\begin{remark}
We make Definition~\ref{def:derivability-base-IPL} substructural to yield Definition~\ref{def:derivability-base-LL} by restricting the assumption to be the atom itself in  \textsc{ref} and by combining assumptions for each collection of premisses in $\textsc{app}$.
\end{remark}

Relative to this notion of derivability in a base, the definition of \emph{support} for ILL without $\oc$ (i.e., IMALL) follows as before:

\begin{defn}[Support for IMALL]
    Support is the smallest relation $\supp{}{}$ satisfying the clauses of Figure~\ref{fig:IMALL}.
\end{defn}

\begin{figure}[t]
{  
\hrule \vspace{1mm}
\[ 
 \begin{array}{l@{\quad}c@{\quad}l@{\quad}r}
            \mbox{$\suppM{\baseB}{\at{S}} \at{p}$} &  \mbox{iff} & \mbox{$\at{S} \deriveBaseM{\baseB} \at{p}$} & \mbox{(At)}  \\[-.5mm]
           \mbox{$\suppM{\baseB}{\at{S}} \phi \multimap \psi$} &  \mbox{iff} & 
            \mbox{$\phi \suppM{\baseB}{\at{S}} \psi$} &   \mbox{($\multimap$)} \\ 
             \mbox{$\suppM{\baseB}{\at{S}} \phi \otimes \psi$} & \mbox{iff} & \mbox{$\forall \baseX \baseGeq \baseB$, $\forall \at{U}$, $\forall \at{p}$, if $\phi \mcomma \psi \suppM{\baseX}{\at{U}} \at{p}$, then $\suppM{\baseB}{\at{S}\mmcomma \at{U}} \at{p}$} & \mbox{($\otimes$)} \\[-.5mm]
             \mbox{$\suppM{\baseB}{\at{S}} \phi \oplus \psi$} & \mbox{iff} & \mbox{$\forall \baseX \supseteq\baseB, \, \forall U, \, \forall p$, if $ \phi\suppM{\baseX}{U}p$ and $\psi\suppM{\baseX}{U}p$, then $\suppM{\baseX}{S \mmcomma U}p$} & \mbox{($\oplus$)} \\[-.5mm]
             \mbox{$\suppM{\baseB}{\at{S}} \phi \with \psi$} & \mbox{iff} & \mbox{$\suppM{\baseB}{S} \phi$ and $\suppM{\baseB}{S} \psi$} & \mbox{($\&$)} \\
             \mbox{$\suppM{\baseB}{\at{S}} \one$} & \mbox{iff} & \mbox{$\forall \baseX \baseGeq \baseB$, $\forall \at{U}$, $\forall \at{p} \in \At$, if $\suppM{\baseX}{\;\at{U}} \at{p}$, then $\suppM{\baseX}{\;\at{S}\;\mcomma\;  \at{U}} \at{p}$} &  \mbox{($\one$)}  \\
             \mbox{$\suppM{\baseB}{\at{S}} \zero$} & \mbox{iff} & \mbox{$S \deriveBase{\baseB} p$ for any $p \in \At$} &  \mbox{($\zero$)} \\
           \mbox{$\suppM{\baseB}{S} \Gamma \mcomma \Delta$} & \mbox{iff} &  \mbox{$\exists \at{U},\at{V}$ s.t. $\at{S} = (\at{U} \mcomma \at{V})$, $\suppM{\baseB}{\at{U}} \Gamma$, and  $\suppM{\baseB}{V} \Delta$} &  \mbox{(\,\,$\mcomma$\,\,)} 
            \\[-.5mm]
             \hspace{-1.4em} \mbox{$\Gamma \suppM{\baseB}{S} \phi$} & \mbox{iff} & 
        \mbox{$\forall \baseX \baseGeq \baseB$ and $\forall \at{U}$, if $\suppM{\baseX}{\at{U}} \Gamma$, then $\suppM{\baseX}{\;\at{S}\;\mcomma\; \at{U}} \phi$} & \mbox{\myhypertarget{imll-inf}{(Inf)}} \\[-.5mm]
        \hspace{-1.4em} \mbox{$\Gamma \suppM{}{} \phi$} & \mbox{iff} & 
        \mbox{$\Gamma \suppM{\baseB}{\emptyset} \phi$ for any base $\baseB$} &
        \end{array}
\]
\hrule
\caption{Base-extension Semantics for IMALL} \label{fig:IMALL}
}
\end{figure}

\begin{theorem}[Gheorghiu, Gu, and Pym~\cite{GGP2023-IMLL},  Buzoku~\cite{Yll-arxiv}]
    $\Gamma \proves[\emph{IMALL}] \phi$ iff $\Gamma \supp{}{} \phi$.
\end{theorem} 

Soundness and completeness are provided in the same way as for IPL in  Section~\ref{sec:b-es:ipl}; that is, soundness follows by the admissibility of the rules from a natural deduction system for ILL with respect to support, and completeness by `simulating' a natural deduction system for ILL by using the clauses of the semantics. 

\begin{remark} \label{rem:vmm-ll}
 Intuitively, the indexed multiset of atoms denotes available \emph{resources}. The Vending Machine Model discussed in Section~\ref{sec:intro} is easily expressed: 
 \begin{quote}
     Let $\Gamma$ be a theory defining the machine's behaviour (e.g., $\Gamma$ 
 is the multiset of just $1{  \euro} \multimap C$) and let $\base{V}$ be any base supporting that theory (i.e., $\suppM{\base{V}}{\emptyset} \Gamma$), then the situation `one euro will buy one chocolate bar' corresponds to the judgement $ \suppM{\base{V}}{1{  \euro}} C$.
 \end{quote}
This stands in contrast to M-tS of linear logics in which the number-of-uses reading is not clearly expressed.
\end{remark}


What about the exponential, `$\oc$'\,~? Intuitively, its resource reading is that the resource to which it is applied is \emph{unbounded} --- zero or arbitrarily many instances of it may be used. This can be expressed in the presented set-up, but requires a lot of delicate overhead, as 
given in \cite{Yll-arxiv,Yll-bristol}, that is not essential for presenting 
our basic account of resource semantics and its use in modelling distributed systems.

\subsection{The Logic of Bunched Implications} \label{sec:b-es:bi}

In this section, we recall the B-eS for BI by the authors~\cite{GGP2024-BI}. Essentially, BI is the \emph{free} combination of IPL --- with connective $\top, \land, \lor, \to$ --- and IMLL --- with connective $\mtop, \mand, \wand$. Intuitively, therefore, the B-eS is a free combination of the semantics for IPL in Section~\ref{sec:b-es:ipl} and for IMLL in Section~\ref{sec:b-es:ll}. While this intuition is fulfilled, there are several subtle technical details that require careful handling. 

A principal challenge of BI is that, as a consequence of having two conjunctions and two (primitive) implications, contexts are not one of the standard data structures (e.g., sets, multiset, lists), but rather \emph{bunches},  a term that derives from the relevance logic literature --- see, for example, Read~\cite{Read1988}. The subtlety is the algebraic properties of these bunches: in BI, part of the bunch admits the structural rules of weakening and contraction, and other parts do not. The B-eS of BI has to precisely express these algebraic properties. 
Furthermore, while our previous work has primarily utilized the `natural deduction' proof formalism, BI is more suitably expressed in sequent calculus form (cf.~\cite{ribbon}), necessitating adjustments to the entire framework of atomic systems.

Notwithstanding the challenges posed by bunches, BI is a well-established logic in systems modelling, as discussed in Section~\ref{sec:intro}. In Section~\ref{sec:res-int}, we see that the treatment of bunches in the B-eS facilitates the expression of certain important features of distributed systems. We now proceed by providing the technical background for BI and defining its B-eS. Recall our previous specification of notation (in Section~\ref{sec:b-es:ipl}). 

Essentially, \emph{bunches} in BI are the free combination of the context constructors of IPL --- denoted $\aunit$ and $\aacomma$ --- and IMLL --- denoted $\munit$ and $\mmcomma$\,. 

\begin{defn}[Bunch]
\emph{Bunches} over some set $\set{X}$ are defined as:
\[  
X ::= x \in \set{X} \mid \aunit \mid \munit \mid X \acomma X \mid X \mcomma X
\]
The set of all bunches over $\set{X}$ is $\Bunches(\set{X})$
\end{defn}

If $\Formulas$ is the set of formulas (over atoms $\At$) in BI, then $\Bunches(\Formulas)$ is the set of bunches of formulas. A \emph{sequent} in BI is a pair $\Gamma \seq \phi$ in which $\Gamma \in \Bunches(\Formulas)$ and $\phi \in \Formulas$. 

\begin{defn}[Sub-bunch]
    If a bunch $\Delta$ is a sub-tree of a bunch $\Gamma$, then $\Delta$ is a \emph{sub-bunch} of $\Gamma$. 
\end{defn}
\begin{remark}
Importantly, sub-bunches are sensitive to occurrence --- for example, in the bunch $\Delta \mcomma \Delta$, each occurrence of $\Delta$ is a different sub-bunch. 
\end{remark}
We may write $\Gamma(\Delta)$ to express that $\Delta$ is a sub-bunch of $\Gamma$. This notation is traditional for \BI{}~\cite{o1999logic}, but requires careful handling. To avoid confusion, we never write $\Gamma$ and later $\Gamma(\Delta)$ to denote the same bunch, but rather maintain one presentation.  The substitution of $\Delta$ for $\Delta'$ in $\Gamma$ is denoted $\Gamma(\Delta)[\Delta \mapsto \Delta']$. 
This is not a universal substitution, but a substitution of the particular occurrence of $\Delta$ meant in writing $\Gamma(\Delta)$. The bunch $\Gamma(\Delta)[\Delta \mapsto \Delta']$ may be denoted $\Gamma(\Delta')$ when no confusion arises.

Since bunches are perhaps not as standardly known as other data structures, we give some detail about their meta-theory that is essential to understanding BI. We require the notion of a \emph{contextual bunch}, which we think of a contextual bunch as a \emph{bunch with a hole} --- cf. \emph{nests with a hole} in work by Br\"unnler~\cite{Brunnler2009}. This is already suggested by the substitution notation, but it is worth giving an explicit mathematical treatment to avoid confusion. 

\labelandtag{cl:BI-BeS-supp:At}{\text{(At)}}
\labelandtag{cl:BI-BeS-supp:conjunction}{\ensuremath{(\land)}}
 \labelandtag{cl:BI-BeS-supp:disjunction}{\ensuremath{(\lor)}}
\labelandtag{cl:BI-BeS-supp:mand}{\ensuremath{(\mand)}}
\labelandtag{cl:BI-BeS-supp:implication}{\ensuremath{(\to)}}
\labelandtag{cl:BI-BeS-supp:wand}{\ensuremath{(\wand)}}
\labelandtag{cl:BI-BeS-supp:bot}{\ensuremath{(\bot)}}
\labelandtag{cl:BI-BeS-supp:top}{\ensuremath{(\top)}}
\labelandtag{cl:BI-BeS-supp:mtop}{\ensuremath{(\mtop)}}
\labelandtag{cl:BI-BeS-supp:inf}{\text{(Inf)}}
\labelandtag{cl:BI-BeS-supp:munit}{\ensuremath{(\munit)}} 
\labelandtag{cl:BI-BeS-supp:aunit}{\ensuremath{(\aunit)}}
\labelandtag{cl:BI-BeS-supp:mcomma}{\ensuremath{(\mmcomma)}} 
\labelandtag{cl:BI-BeS-supp:acomma}{\ensuremath{(\aacomma)}}

\begin{figure}[t]
\hrule \vspace{1mm}
\[
 \begin{array}{lc@{\quad}lr}
        \suppBI{\at{S}}{\base{B}} \at{p}  
        & \text{iff} & \at{S} \proves[\base{B}]\at{p} &
        \ref{cl:BI-BeS-supp:At} \\[-.5mm]
        \suppBI{\at{S}}{\base{B}} \phi \land \psi 
        & \text{iff} & \mbox{$\forall \base{X} \baseGeq \base{B}, \forall \at{U}(\cdot) \in \BunchesWithHole(\At), \forall p \in \At$, if $\phi \addcontext \psi\suppBI{\at{U}(\cdot)}{\base{X}} \at{p}$, then $\suppBI{\at{U(S)}}{\base{X}} \at{p}$}
       & \ref{cl:BI-BeS-supp:conjunction} \\[-.5mm]
        \suppBI{\at{S}}{\base{B}} \phi \mand \psi 
        & \text{iff} & \mbox{$\forall \base{X} \baseGeq \base{B}, \forall \at{U}(\cdot) \in \BunchesWithHole(\At), \forall p \in \At$, if $\phi \multcontext \psi\suppBI{\at{U}(\cdot)}{\base{X}} \at{p}$, then $\suppBI{\at{U(S)}}{\base{X}} \at{p}$} &
        \ref{cl:BI-BeS-supp:mand} \\[-.5mm]
        \suppBI{\at{S}}{\base{B}} \phi \lor \psi  
        & \text{iff} & \forall \base{X} \baseGeq \base{B}, \forall \at{U}(\cdot) \in \BunchesWithHole(\At), \forall p \in \At, \text{if } \phi \suppBI{\at{U}(\cdot)}{\base{X}} \at{p} \text{ and } \psi \suppBI{\at{U(\cdot)}}{\base{X}} \at{p}, \text{ then}  \suppBI{\at{U(S)}}{\base{X}} \at{p} & \ref{cl:BI-BeS-supp:disjunction} \\[-.5mm]
        \suppBI{\at{S}}{\base{B}} \phi \to \psi 
        & \text{iff} &
        \phi \suppBI{\at{S} \aacomma (\cdot)}{\base{B}} \psi & \ref{cl:BI-BeS-supp:implication} \\[-.5mm]
        \suppBI{\at{S}}{\base{B}} \phi \wand \psi 
        & \text{iff} & \phi \suppBI{\at{S} \mmcomma(\cdot)}{\base{B}} \psi &
        \ref{cl:BI-BeS-supp:wand} \\[-.5mm]
        
        \suppBI{\at{S}}{\base{B}} \bot & \mbox{iff} &  \forall \at{U}(\cdot) \in \BunchesWithHole(\At), \forall \at{p} \in \At, \suppBI{\at{U(S)}}{\baseB} \at{p}  & \ref{cl:BI-BeS-supp:bot} \\[-.5mm]
        \suppBI{\at{S}}{\base{B}} \top & \text{iff} & \forall \base{X} \baseGeq \base{B}, \forall \at{U(\cdot)} \in \BunchesWithHole(\At), \forall \at{p} \in \At, \text{ if } \suppBI{\at{U(\aunit)}}{\baseX} \at{p}, \text{ then} \suppBI{\at{U(S)}}{\base{X}} \at{p} & \ref{cl:BI-BeS-supp:top} \\[-.5mm]
        \suppBI{\at{S}}{\base{B}} \mtop 
        & \text{iff} &\forall \base{X} \baseGeq \base{B}, \forall \at{U(\cdot)} \in \BunchesWithHole(\At), \forall \at{p} \in \At, \text{ if } \suppBI{\at{U(\munit)}}{\baseX} \at{p}, \text{ then} \suppBI{\at{U(S)}}{\base{X}} \at{p} & \ref{cl:BI-BeS-supp:mtop} \\
        %
    %
     \suppBI{\at{S}}{\base{B}} \Gamma \mcomma \Delta 
    & \mbox{iff} & \mbox{$\exists \at{Q_1}, \at{Q_2} \in \Bunches(\At)$ such that $\at{S} \bunchStrongerThan \at{Q_1} \mcomma \at{Q_2}$, $\suppBI{\at{Q_1}}{\baseB} \Gamma$, and  $\suppBI{\at{Q_2}}{\baseB} \Delta$}  & \ref{cl:BI-BeS-supp:mcomma} \\[-.5mm]
    %
    \suppBI{\at{S}}{\baseB} \Gamma \acomma \Delta  
    & \mbox{iff} & \mbox{$\exists \at{Q_1}, \at{Q_2} \in \Bunches(\At)$ such that $\at{S} \bunchStrongerThan \at{Q_1}$, $\at{S} \bunchStrongerThan \at{Q_2}$, $\suppBI{\at{Q_1}}{\baseB} \Gamma$, and  $\suppBI{\at{Q_2}}{\baseB} \Delta$} 
    & \ref{cl:BI-BeS-supp:acomma}\\
            \hspace{-1em} \Gamma \suppBI{\at{R(\cdot)}}{\base{B}} \psi 
        & \text{iff} &\forall \base{X} \baseGeq \base{B}, \forall \at{U} \in \Bunches(\At), \text{ if } \suppBI{\at{U}}{\baseX} \Gamma, \text{ then } \suppBI{\at{R(U)}}{\baseX} \psi 
       & \ref{cl:BI-BeS-supp:inf} \\
      \hspace{-1em}  \Gamma \suppBI{}{} \phi & \mbox{iff} & \mbox{$\Gamma \suppBI{(\cdot)}{\baseB} \psi$ for any base $\baseB$}  
    \end{array}
\]
\hrule
\caption{Base-extension Semantics for BI} \label{fig:BI}
\end{figure}


\begin{defn}[Contextual Bunch]
A \emph{contextual bunch} (over $\set{X}$) is a function $b:\Bunches(\set{X}) \to \Bunches(\set{X})$ such that there is $\Gamma(\Delta) \in \Bunches(\set{X})$ and $b(\varSigma) = \Gamma(\varSigma)$ for any $\varSigma \in \Bunches(\set{X})$. The identity on $\Bunches(\set{X})$ is a contextual bunch, it is denoted $(\cdot)$. The set of all contextual bunches (over $\set{X}$) is $\dot{\Bunches}(\set{X})$.
\end{defn}

Observe that $\dot{\Bunches}(\set{X})$ can be identified as the subset of $\Bunches(\set{X}\cup\{\circ\})$, where $\circ \not \in \set{X}$, in which bunches contain a single occurrence of $\circ$. Specifically,  if $b(x) =\Gamma(x)$, then identify $b$ with $\Gamma(\circ) \in \Bunches(\set{X}\cup\{\circ\})$. We write $\Gamma(\cdot)$ for the contextual bunch identified with $\Gamma(\circ)$. 
We require a notion of equivalence on bunches. 

\begin{defn}[Coherent Equivalence]
Bunches $\Gamma,\Gamma' \in \Bunches(\set{X})$ are \emph{coherently equivalent} when $\Gamma \equiv \Gamma'$, where $\equiv$ is the least relation satisfying: commutative monoid equations for $\fatsemi$ with unit $\aunit$ and for $\fatcomma$ with unit $\munit$, and  congruence (i.e., if $\Delta\equiv\Delta' $ then $\Gamma(\Delta) \equiv \Gamma(\Delta'))$.
 \end{defn}

In practice, bunches are regarded as the syntactic constructions in $\Bunches(\mathbb{X})$ modulo coherent equivalence.

We require a way to express the structurality of the additive context-former. 

\begin{defn}[Bunch-extension]
 The \emph{bunch-extension} relation $\bunchStrongerThan$ is the least relation satisfying: 
    \begin{itemize}[nosep,label=--] 
        \item (\textsc{weakening}) if $\Gamma \equiv \Gamma'[\Delta \mapsto \Delta \fatsemi \Delta']$, then $\Gamma \bunchStrongerThan \Gamma'$
        \item (\textsc{transitivity}) if $\Gamma \bunchStrongerThan \Gamma'$ and $\Gamma' \bunchStrongerThan\Gamma''$, then $\Gamma \bunchStrongerThan\Gamma''$.
    \end{itemize}
\end{defn}
Note that (\textsc{reflexivity}) $\Gamma \bunchStrongerThan \Gamma$ is derivable here. 

\emph{Base-extension Semantics.} The notion of \emph{bases} used for BI diverges from that of Section~\ref{sec:b-es:ipl} and Section~\ref{sec:b-es:ll} as the natural deduction format is insufficiently expressive to make all the requisite distinctions about multiplicative and additive structures in bunches. Accordingly, we move to a sequent calculus format. 

An \emph{atomic sequent} is a pair $P \seq p$ where $P \in \Bunches(\At)$, $p \in \At$. An atomic rule is a rule over atomic sequents,
\[ {
\infer[\rn{A}]{\at{P} \seq \at{p}}{} \qquad 
       \infer[\rn{R}]{\at{P} \seq \at{p}}{\at{P}_1 \seq \at{p}_1 & \hdots & \at{P}_n \seq \at{p}_n} }
       \]
A base is a set of atomic rules; we use the same same conventions as in Section~\ref{sec:b-es:ipl} and Section~\ref{sec:b-es:ll}. They behave as sequent calculus systems.

\begin{defn}[Derivability in a Base] \label{def:derivability-base-BI}
Derivability in a base $\baseB$ is the smallest relation $\proves{\baseB}$ satisfying the following:
\begin{itemize}[nosep,label=--]
    \item \myhypertarget{def:derivability-in-a-base:taut}{\textsc{taut}}: \mbox{$\at{p} \proves[\baseB] \at{p}$, for all $p\in\At$} 
    \item \myhypertarget{def:derivability-in-a-base:initial}{\textsc{initial}}: \mbox{If $\rn{A} \in \baseB$, then $\at{P} \proves[\baseB] \at{p}$} \item \myhypertarget{def:derivability-in-a-base:rule}{\textsc{rule}}: \mbox{If $\rn{R} \in \baseB$ and $\at{P}_1 \proves[\baseB] \at{p}_1, \dots, \at{P}_n \proves[\baseB] \at{p}_n$, then $\at{P} \proves[\baseB] \at{p}$} 
    \item \myhypertarget{def:derivability-in-a-base:weak}{\textsc{weak}}:  \mbox{If $\at{P}(\at{Q}) \proves[\baseB] \at{p}$, then $\at{P}(\at{Q} \fatsemi \at{Q}') \proves[\baseB] \at{p}$, for any $\at{Q}' \in \Bunches(\Atoms)$} 
    \item \myhypertarget{def:derivability-in-a-base:cont}{\textsc{cont}}:  \mbox{If $\at{P}(\at{Q}\fatsemi \at{Q}) \proves[\baseB] \at{p}$, then $\at{P}(\at{Q}) \proves[\baseB] \at{p}$} 
\item \myhypertarget{def:derivability-in-a-base:exch}{\textsc{exch}}:  \mbox{If $\at{P}(\at{Q}) \proves[\baseB] \at{p}$ and $\at{Q} \equiv \at{Q}'$, then $\at{P}(\at{Q}') \proves[\baseB] \at{p}$}
\item \myhypertarget{def:derivability-in-a-base:cut}{\textsc{cut}}:  \mbox{If $\at{T} \proves[\baseB] \at{q}$ and $\at{S}(\at{q}) \proves[\baseB] \at{p}$, then $\at{S}(\at{T}) \proves[\baseB] \at{p}$.}
\end{itemize}
\end{defn}

\begin{remark} \label{rem:nd-seq}
    While Definition~\ref{def:derivability-base-BI} appears more complex than its analogues for IPL (Definition~\ref{def:derivability-base-IPL}) and LL (Definition~\ref{def:derivability-base-LL}), this is caused by the change of formalism; in particular,  \textsc{taut} corresponds to \textsc{ref},  \textsc{initial} to \textsc{app}$_1$,  \textsc{rule} to \textsc{app}$_2$, and \textsc{cut} to the composition natural deduction proofs. The remaining conditions --- namely, \textsc{weak}, \textsc{cont}, and \textsc{exch} --- simply express the algebraic properties of bunches.
\end{remark}

\begin{defn}[Support for BI]
    Support is the smallest relation $\supp{}$ satisfying the clauses in Figure~\ref{fig:BI}, in which $S \in \Bunches(\At)$, $\at{R}(\cdot) \in \BunchesWithHole(\At)$, $\Gamma, \Delta \in \Bunches(\Formulas)$.
\end{defn}

\begin{theorem}[Gu, Gheorghiu, and Pym~\cite{GGP2024-BI}]
    $\Gamma \proves[\emph{BI}] \phi$ iff $\Gamma \supp{}{} \phi$.
\end{theorem} 
The proof of this again follows the techniques used by Sandqvist~\cite{Sandqvist2015base}; in particular, completeness follows from simulating a natural deduction proof systems for BI expressed in sequent calculus form (see, e.g., 
\cite{o1999logic,Pym2002book,Brotherston2012,HarlandPym2003,Galmiche2005,GGP2024-BI}). Moreover, using the correspondences in Remark~\ref{rem:nd-seq}, it is straightforward to express the B-eS of IPL and IMLL in Section~\ref{sec:b-es:ll} as a restriction of the B-eS of BI herein. Unlike in Section~\ref{sec:b-es:ll}, in the B-eS of BI presented here, these fragments are freely combined. 

There are a few things to remark upon about the support relation.

\begin{remark}
In contrast to the treatment of IPL~\cite{Sandqvist2015base} discussed in Section~\ref{sec:intro}, support is now also parametrized by a bunch of atoms. In Section~\ref{sec:res-int}, we see that these may usefully be thought of as \emph{resources}. The role of a contextual bunch can be seen particularly clearly in the \ref{cl:BI-BeS-supp:inf} clause in Figure~\ref{fig:BI}: the resources required for the sequent $\Gamma \seq \phi$ are combined with those required for $\Gamma$ in order to deliver those required for $\phi$. 
\end{remark}

\begin{remark}
The treatment of the support of a combination of contexts $\Gamma \mcomma \Delta$ follows the na\"{\i}ve Kripke-style interpretation of multiplicative conjunction, corresponding to an introduction rule in natural  deduction, but the support of the tensor product $\phi \otimes\psi$ follows the form of a natural deduction elimination rule. The significance of this is explored in Section~\ref{sec:res-int}. 
\end{remark}

\begin{remark}
    The Vending Machine Model is expressed exactly as in Section~\ref{sec:b-es:ll} (Remark~\ref{rem:vmm-ll}). This stands in stark contrast to the situation in Section~\ref{sec:intro} where the extant resource semantics of the logics require fundamentally different approaches to model the same system.
\end{remark}

We conclude the technical background required for this paper: we have provided inferentialist semantics for IPL, IMALL, and BI. It remains to give a resource interpretation of these semantics. 

\section{Inferentialist Resource Semantics}  \label{sec:res-int} 


\noindent 
We give a systematic account of how B-eS can be used to 
give an account of the resource semantics in which formulae are 
interpreted as assertions about the flow of resources in a distributed system, expressing both the sharing/separation interpretation of BI \cite{pym2019resource} and the number-of-uses reading of LL. It is also related to the syntactic resource semantics of Pfenning and Reed \cite{Reed2009}.  
After the conceptual understanding in this section, we then give detailed examples in Section~\ref{sec:example}.

In the context of models of distributed systems that are formulated in terms of locations, resources, and processes, we begin with a conceptual definition of \emph{resource semantics}, as follows: 
\begin{quote}
    \emph{A resource semantics for a system of logic is an interpretation of its formulae as assertions about states of processes and is expressed in terms of the resources that are manipulated by those processes}.  
\end{quote}
This definition requires a few notes: we intend no restriction on the assertions that are to be included in the scope of this definition;
assertions may refer not only to ground states but also `higher-order' assertions about state transitions. Moreover, we intend the manipulation of resources by the system's processes to include, \emph{inter alia}, consuming/using resources, creating resources, copying/deleting resources, and moving resources between locations.
Furthermore, we require that an adequate resource semantics should be able to provide accounts of the following concepts:  
counting of resources, composition of resources, comparison of resources, sharing of resources, and separation of resources. 

A system's policy is what determines the system's processes and how they manipulate resources --- see Section~\ref{sec:intro}. In an \emph{inferential} account of resource semantics, those behaviours determine the meaning of the system --- see Section~\ref{sec:b-es}. We now show how B-eS is used to model policies directly, and then illustrate this account through examples in Section~\ref{sec:example}.




In the B-eS we have presented in Section~\ref{sec:b-es}, it is the (Inf) clause that articulates the consequence relation free of the structure of the connectives. Indeed, 
whilst the semantic clauses for connectives vary in the B-eS for LL and BI, their (Inf) clauses follow a similar rationale. This is no coincidence: $\Gamma \suppM{}{} \phi$ is about the transmission of validity, thus its definition reflects the semantic paradigm rather than of the specific logic system.
Indeed, analogously in M-tS, $\Gamma \suppM{}{} \phi$ expresses that `for any model $\mathfrak{M}$, if $\Gamma$ is true in $\mathfrak{M}$, then so is $\phi$', regardless of the specific logic and semantics of interest. The power of B-eS to unify the aforementioned resource interpretations can be summarized in the following general pattern of support judgement and its semantics clauses for LL 
and BI --- (Inf) is the key clause that connects the left- and right-sides of the support judgement, free of reference to connectives:
\[
\label{eq:general-inf}
    \mbox{$\Gamma \suppM{\baseB}{S(\cdot)} \phi 
\quad$ iff $\quad
    \forall \baseC \baseGeq \baseB, \forall U \in \GenRes(\At)$, if $\suppM{\baseC}{U} \Gamma$, then $\suppM{\baseC}{S(U)} \phi$} 
    \tag{Gen-Inf} 
\]

Let us spell out each ingredient of \eqref{eq:general-inf} and their roles in the resource semantics, which describes the execution of certain policy in a system:
\begin{itemize}[nosep,label=--]
    \item $\phi$ is a formula of the chosen logic. It is an assertion describing (a possible state of) the system
    \item $\Gamma$ is a collection (e.g., multiset, bunch, etc.) of formulas. It specifies a policy describing the 
    executions of a system's processes 
    \item $\GenRes(\At)$ is some choice of atomic resource (e.g., multisets of atoms in ILL, bunches of atoms in BI)
    \item $S(\cdot)$ is some contextual atomic resource, such that whenever combined with some atomic resource in $\GenRes(\At)$, it returns a `richer' atomic resource (e.g., multisets of atoms in IMLL, contextual bunches in BI). It specifies the resources that are available for the system model's processes to execute according to the given policy
    \item $\baseB, \baseC$ are bases of one's choice. They are models, and $\suppM{\baseC}{U} \Gamma$ reads as $\baseC$ is a model of policy $\Gamma$ when supplied with resource $U$.
\end{itemize}
Putting all these together, the support judgement $\Gamma \suppM{\baseB}{S(\cdot)} \phi$ says that, if policy $\Gamma$ were to be executed with contextual resource $S(\cdot)$ based on the model $\baseB$, then the result state would satisfy $\phi$. \eqref{eq:general-inf} explains how such execution is triggered: for arbitrary model $\baseC$ that extends $\baseB$, if policy $\Gamma$ is met in model $\baseC$ using some resource $U$, then $\Gamma$ could be executed, and the resulting state --- which consumes $U$ in the available contextual resource $S(\cdot)$ --- satisfies $\phi$.

We demonstrate how this general resource interpretation retrieves the number-of-uses reading and the sharing/separation semantics when instantiated in ILL (in particular, IMALL) and BI, respectively.

\subsection{Linear Logic}

Following Section~\ref{sec:b-es:ll}, we restrict ourselves to IMALL.
In this context, a multiset of atoms $P$ denotes a contextual resource $S(\cdot)$ through multiset union --- that is, $S(\cdot): U \mapsto P \mcomma U$. 

The number-of-uses interpretation of LL instantiates the definition of resource semantics above as follows: a formula $\phi$ asserts that the system has some resources \quotes{\phi} (discussed below), and an implication $\phi \multimap \psi$ asserts that the system has a process that uses the resources \quotes{\phi} and yields resources  \quotes{\psi}. What resources are \quotes{\phi} and \quotes{\psi}? That depends on the policy $\baseB$ governing the system; for example, when the policy is empty ($\baseB = \emptyset$) and $\phi$ and $\psi$ are atoms $p$ and $q$, respectively, then \quotes{\phi} is $p$ and $\quotes{\psi}$ is $q$ --- we give further examples below. Understanding that this interpretation is relativized to a policy is essential; for example, when $\baseB$ contains an axiom $\rn{A}$, then one has an indefinite amount of the resource $c$ available. Importantly, this handling of resource is the idea driving the `simulation' proof of the completeness of the semantics --- see Gheorghiu et al.~\cite{GGP2023-IMLL} --- discussed in Section~\ref{sec:b-es:ll}.

The clauses in Figure~\ref{fig:IMALL} enable this way of reading number-of-uses as a resource semantics; that is, they explicate inductively how a formula is interpreted as a collection of resources relative to a policy --- from $\suppM{\baseB}{P} \phi \multimap \psi$, one recovers a judgement $\suppM{\baseB \cup \baseC}{P \mmcomma U} \psi$, in which $U$ is \quotes{\phi} as determined by $\base{C}$. We illustrate for some of connectives to show how their number-of-uses readings manifest: 
\begin{itemize}[nosep,label=--]
   \item $\&$: This clause very clearly says that $\phi \with \psi$ denotes that $\phi$ and $\psi$ are available processes that each require the same resources; in other words, both process $\phi$ and $\psi$ \emph{may} be done, but only one \emph{will} be.
   \item $\mmcomma$\,: Similar to the above, both process $\phi$ and $\psi$ may be done, but this time both \emph{must} be done, so the resources must be divided in some suitable way. 
    \item $\otimes$: One may expect this clause to take precisely the same form as (\,$\mcomma$\,), but taking the elimination form instead ensures a certain coherence in the system~\cite{GGP2023-IMLL}. Indeed, taking this form, the clause allows us to see precisely how a formula denotes a collection of resources relative to a policy. 
    
    \quad Let us return to $\phi \multimap \psi$. We already considered the case in which both $\phi$ and $\psi$ are atoms. Suppose now that $\phi$ is a \emph{tensor} of atoms (i.e., $\phi:=p_1 \otimes \ldots \otimes p_n$). Again, choosing the simplest possible $\baseB$ to model this process and make \quotes{\phi} as plain as possible, the judgement $\supp{}{} \phi \multimap \psi$ can be reduced to $p_1\mcomma\ldots\mcomma p_n \deriveBase{\baseB} q$, in which $\baseB$ is simply a policy for a process that \emph{uses resources $\phi$ and yields the resources $\psi$}. 
    
    \quad In general, one has $\quotes{\phi}\deriveBase{\baseB \cup \base{C}} q$, in which $\base{C}$ ensures that $\quotes{\phi}$ \emph{behaves} as $p_1 \otimes \cdots \otimes p_n$. 
    For example, $\quotes{\phi}$ can be a single atom and $\baseC$ may simulate the appropriate introduction and elimination rules,     
    \[ {
    \infer{\quotes{\phi}}{\{p_1\} & \ldots & \{p_n\}} \qquad \infer{x}{\raisebox{-1em}{$\{\quotes{\phi}\}$} & \left\{
    \begin{array}{c}
    [p_1 \mcomma \ldots \mcomma p_n] \\
    x 
    \end{array}
    \right\}} }
    \]
    \normalsize
    A similar analysis can be made for when $\psi$ is a tensor-formula.  
    \item $\oplus$: This connective denotes non-deterministic choice: one of $\phi$ or $\psi$ will fire, but we do not know which. The elimination-rule format of the clause expresses precisely this: what the process $\phi$  and $\psi$ can both yields with resources $U$,  is yielded by $\phi \oplus \psi$ with resources $U$.  
\end{itemize}

This suffices to illustrate how the number-of-uses reading of LL manifests in its B-eS; the remaining connectives follow similarly. 
In the treatment of $\otimes$, we have concentrated on it as a pre-/post- condition of $\multimap$ (as opposed to the combination of two process that both execute as in $\mmcomma$\,). This is to explicate how the format of the clause, which indeed passes through $\mmcomma$\,, supports the interpretation of formulae as assertions about resources. 

\subsection{Bunched Implications} \label{subsec:BI}

Briefly, a distributed system is comprised of components that exchange and process resources (see Section~\ref{sec:generality} for more on this). 
Two components \emph{share} resources when they use the available resource at the same time and \emph{separate} resources when the available resources must be divided between components. The sharing/separation interpretation of BI (see Pym~\cite{pym2019resource}) pertains to modelling this aspect of the architecture of distributed systems. We illustrate how this manifests in BI's B-eS relative to the notation in (Gen-Inf).

First, the bunch of atoms $U$ denotes the available resources. If $V$ and $W$ are collections of resources, then $V \aacomma W$ denotes a collection of resources such that each sub-collection may be shared --- that is, components that share resources may use both $V$ and $W$ simultaneously, just $V$, just $W$, or one component uses $V$ while the other uses $W$ --- and $V \mcomma W$ denotes a collection of resource such that each sub-collection must be used separately. This follows from the $\aacomma$- and $\mmcomma$-clauses, observing also the role played by bunch-extension in those clauses. 

Second, the bunch $\Gamma$ describes the architecture of the system. The two context-formers provide the sharing/separation reading: if $\Gamma$ and $\Delta$ describe system policies for two components, then $\Gamma \acomma \Delta$ describes a system policy in which those components share resources, and $\Gamma \mcomma \Delta$ describes a system policy in which they separate resources. Again, this is immediately expressed by the $\aacomma$- and $\mmcomma$-clauses as the available resources are copied or divided between the sub-bunches, respectively. Intuitive readings of the associated units follow similarly. 

Third, 
$\phi$ is an assertion about the state of the system in terms of its sharing/separation architecture; in particular, considering some of the connectives,  
\begin{itemize}[nosep,label=--]
    \item $\mand$: denotes the combined effect of two separating components of the system taken as a single component
    \item $\land$: denotes the combined effect of two sharing components of the system taken as a single component
    \item $\wand$: denotes that, if the state of the system satisfies the antecent assertion, then the system can be modified to satisfy the consequent assertion
    \item $\to$: denotes that, if the state of the system satisfies the antecedent, then the (unmodified) system 
    also satisfies the consequent. 
\end{itemize}
Presented individually in this way, the significance of these readings is, perhaps, obscured. In the base case, an atom $p$ may be thought of as the assertion that the system has a certain resource; thus, a formula $p \wand (q_1 \mand q_2)$ asserts that the system can move from a state in which a component may use a resource $p$ to a state in which two separate components may use $q_1$ and $q_2$. This is made more concrete in Section~\ref{sec:example} with examples given in which these readings are instantiated in the setting of a relatable distributed systems. 


Observe that restricting to the multiplicative fragment of BI one also recovers IMLL, hence one has the number-of-uses interpretation described for LL above. However, this reading actually extends to the whole of BI. The sharing/separation reading 
arises from the interpretation of the meta-connectives (i.e., the context-formers and their units), but one can consistently keep a number-of-uses interpretation of formulae as the way in which they make assertions about the system (i.e., they express how resources behave within the system).

The implications assert that there is a \emph{process} that transitions the system from a state satisfying the antecedent to one satisfying the consequent; they are distinguished by whether or not the processes \emph{modify the system} when executed. What do we mean by `modify'? In the basic case, we simply mean the consuming of resources; for example, both $p \wand q$ and $p \to q$ refers to a process in which a system moves from a state in which it has a resource $p$ to one in which it has a $q$, but the former does it while consuming $p$ (in the sense of the number-of-uses reading) and the latter only requires that $p$ is available. In the general case, we must account for the fact that the precondition of the process may itself be a process; for example, $(\phi \wand \psi) \wand \chi$ denotes a process that modifies the system from the state $\phi \wand \psi$ (i.e., `there is a process \ldots') to the state $\chi$.

Note, since $\wand$ modifies the system, its required resources must be private (not shared with other processes) so it uses separated resources. Since $\to$ does not modify the system its required resources are the kind that may be shared. 

\subsection{Thesis} \label{subsec:thesis}

We have described how a general view of B-eS can be instantiated to explicate the resource semantics both of logics that exhibit the sharing/separation semantics of propositions and logics that support the number-of-uses readings of propositions. 

Recall from Section~\ref{sec:intro} and our discussion of (\ref{eq:general-inf}) that an inferentialist account of resource semantics is one in which policies determine behaviours and these behaviours collectively give the meaning of the system. Thus we conclude this section by asserting the thesis of this paper:
\begin{quote}
\emph{The paradigm of base-extension semantics provides an inferentialist account of resource semantics that uniformly encompasses both the number-of-uses readings --- as found in the family of linear logics --- and the sharing/separation semantics --- as found in bunched logics, such as BI and relevance logics.} 
\end{quote}
Observe that this uniformity stands in contrast to extant accounts of these readings of linear and bunches logic that proceed through different frameworks --- namely, proof theory and model theory.

We now illustrate this thesis, in Section~\ref{sec:example}, with two familiar yet evidently generic examples of distributed systems and, in Section~\ref{sec:generality}, we describe the 
realization of the thesis for distributed systems more generally.

\section{Generic Examples: Airport Security and MFA}  \label{sec:example} 
Having developed (a sketch of) an inferentialist 
resource semantics, we now illustrate the ideas by exploring some substantial examples. These examples are intended to be familiar and relatable settings in which policies are applied to located resources (e.g.,~\cite{CMP2012,CP15,AP16}). Despite being specific in their details, they are both structurally and conceptually quite generic.
Since the resource semantics of BI provided by the B-eS expresses both sharing/separation and number-of-uses 
in systems modelling, we concentrate on this logic. 

\subsection{Airport Security Processes} \label{subsec:airport}

\begin{figure}[t]
\hrule
    \centering
    \includegraphics[width=0.8\linewidth]{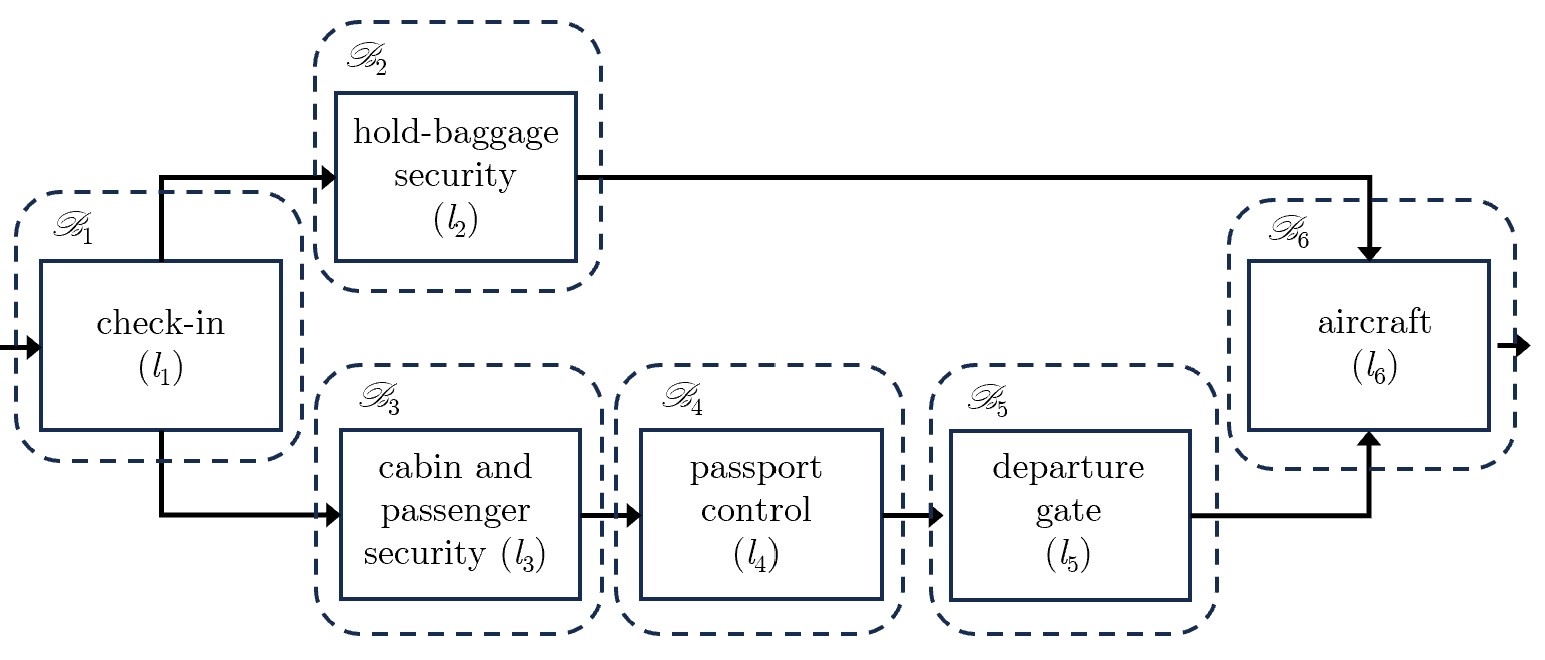}
    \hrule
    \caption{The Airport Security Architecture} 
    \label{fig:airport}
\end{figure}

A model of the departure security process at an airport shown in Figure~\ref{fig:airport}. There are six locations --- $l_1$, \ldots, $l_6$ --- each of which has an associated \emph{policy}, modelled by a base ($\baseB_i$ at $l_i$). There are lots of natural notions of \emph{resource} in this setting (e.g., machines, baggage, passengers, etc), but to keep things simple and focused on \emph{security}, we shall restrict attention to documentation (i.e., passports, tickets, bar-codes, etc). 

You arrive at \emph{check-in} ($l_1$). You show your passport, receive your boarding-card, and drop your hold-baggage. This situation is described by $\Gamma_1 := p \wand ((p \land t) \mand h)$ --- atom $p$ denotes your passport, $t$ denotes the boarding-card (ticket), and $h$ denotes the 
baggage-label. The $\mand$ is used because the system bifurcates at this point and the resources  $p \land t$ and $h$ go to \emph{separating} components, and the $\wand$ is used because the system is modified: the state of passport $p$ is changed as it only goes down one branch (the same as your ticket, $t$) and is no longer globally available. We explain the use of $\mand$ at $l_1$ in more detail below. An inferential model is given by a base $\baseB_1$ supporting $\Gamma_1$.


Next, two processes occur in parallel, one through $l_2$ and one through $l_3,l_4,l_5$. They are described by $\Delta_1$ and $\Delta_2$, respectively. Since they occur separately --- that is, without sharing resources, this portion of the system is described by the bunch $\Gamma_2 := \Delta_1 \mcomma \Delta_2$. Indeed, the $\mmcomma$ at $l_2$ matches the $\mand$ used in $l_1$ above.

The top path of Figure~\ref{fig:airport} passes through \emph{hold-baggage security} ($l_2$). Here, the label $h$ on your baggage is verified (and the baggage itself is checked). That $h$ validates is denoted by $\Delta_1 := h \wand s_{\text{hold}}$.

The bottom path of Figure~\ref{fig:airport} passes through $l_3$, $l_4$, and $l_5$, modelled by $\chi_1$, $\chi_2$, and $\chi_3$, respectively. This branch does not consume resources, so each is modelled using $\to$ (as opposed to $\wand$). They share resources (e.g., passport), so $\Delta_2:= \chi_1 \fatsemi \chi_2 \fatsemi \chi_3$. Their ordering is enforced by the security tokens (see below), otherwise $\to$ may be used in place of $\fatsemi$\,. We consider each location separately. 
\begin{itemize}[label=--]
\item You arrive at \emph{security} ($l_3$). You must present a valid ticket to pass through the security gates. We denote this situation $\chi_1:=t \to s_{\text{cab}}$. 
\item You enter \emph{passport control} ($l_4$). Your passport is validated and you are granted access to the gates. We denote this situation $\chi_2:=(s_\text{cab} \land p) \to s_\text{pass}$.
\item You arrive at \emph{the gate} ($l_5$). Your passport and ticket are checked and you are granted access. We denote this situation $\chi_3:= (s_\text{pass} \land p\land t) \to s_{\text{gate}}$. 
\end{itemize} 

The two separate processes (i.e., $l_2$ and $l_3$ to $l_5$) come together at \emph{the aircraft} ($l_6$).  Here the ground-crew and the air-crew have  check-lists that need to be processed and combined, but that is all hidden from you.  From your perspective, both you and your hold-baggage must have been authorized to board and then, assuming clearance, the plane takes off ($f$ --- flight). We  model this situation by  $\Gamma_3:=(s_\text{gate} \mand  s_\text{hold}) \wand f$.

We have modelled airport security in three parts, producing $\Gamma_1$, $\Gamma_2$, and $\Gamma_3$. How should they be composed into a single theory describing the whole system at once? Observe that overall Figure~\ref{fig:airport} describes a \emph{single process} in which a passport $p$ yields flight $f$. Hence, it is modelled by an implication. The details of the process are what are described by $\Gamma_1$, $\Gamma_2$ ($\Delta_1$ and $\Delta_2$), and $\Gamma_3$, so we take their formula translations $\phi_1$, $\phi_2$ ($\psi_1$ and $\psi_2$), and $\phi_3$, respectively (i.e., replace $\aacomma$ with $\land$, and $\mmcomma$ with $\mand$); that is, 
\[ {
\Gamma:=\phi_1 \wand ( (\psi_1 \mand \psi_2 ) \wand \phi_3 )
} \]
The $\wand$ (rather than $\to$) is appropriate because the system only flows one way; for example, having passed $l_1$, one cannot return to it later, the system has changed. What process do the implications explicit above represent? They are the \emph{interfacing} between the various sections of system; this is part of the system and are modelled by $\baseC_1$ and $\baseC_2$, respectively. 

What $\baseB$ models the policy described by $\Gamma$? The compositional approach by which we described the model is entirely suitable, $\baseB := \baseB_1 \cup \ldots \cup \baseB_6 \cup \baseC_1 \cup \baseC_2$ suffices. This describes how the overall system is modelled, but it is instructive to look at bases in more detail to see how modelling of the components in this compositional approach is done.

We require $\baseB_1$ to support the formula $p \wand (h \mand (t \land p))$; that is, we require $\suppBI{\munit}{\baseB_1} p \wand ((p \land t) \mand h)$ to hold. Recall the clause for $\mand$,
\[ {  
\suppBI{p}{\base{B}_1} (p \land t) \mand h \quad \mbox{iff} \quad \mbox{$\forall \baseX \supseteq \baseB_1 \, \forall x \in \At$, if $(p \aacomma t) \mcomma h \suppBI{U(\cdot)}{\baseX} x$, then  $\suppBI{U(p)}{\baseX} x$}
} \]
This means that in any situation in which one can use the collection of resources $h \mcomma (t \aacomma p)$, it suffices to use the resource $p$. So, in the simplest case, it suffices for the base $\base{B}_1$ to contain the following rules: for any $U(\cdot) \in \BunchesWithHole(\At)$ and $x\in\At$, 
\[
\infer{U(p) \seq x}{U(h\mcomma t)\seq x}
\] 
\normalsize

\noindent This is the coarsest possible approach. In practice, \emph{check-in} is a whole system unto itself (see the discussion on \emph{substitution} in Section~\ref{sec:generality}) and there are many internal processes that run. For example, check-in may proceed as follows: the system extracts from the passport three different parts, the name $p_{\text{name}}$, the date of birth $p_{\text{dob}}$, and the passport-number $p_{\text{num}}$; uses the passport number to issue the ticket; and uses the name on the passport and the ticket together to issue the hold-baggage label. In this case, $\baseB_1$ may have the following rules:
\[ { 
\infer{p \seq p_\text{name}}{} \qquad \infer{p \seq p_\text{dob}}{} \qquad \infer{p \seq p_\text{num}}{} \qquad 
\infer{U(p) \seq x}{p_\text{num} \seq t & p_{\text{name}} \mcomma t \seq h & U(h\mcomma t) \seq x} \qquad \raisebox{1em}{$\ldots$}
} \]
where $\ldots$ denotes the part of the base that models the way in which the passport number $(p_\text{num})$ is used to issue the ticket $(t)$ and the name and ticket together $(p_\text{name} \mcomma t)$ are used to issue the hold-baggage label $(h)$. 
The same kind of flexibility in the modelling occurs at each location. 



\subsection{Multi-factor Authentication (MFA)} 
\label{subsec:mfa}

To illustrate how generic the notion of system used really is, we give an immediate translation to the setting of information security. The goal of MFA is to increase the overall security of systems by requiring multiple forms of verification before granting access; for example, login systems on a banking app. Such verification forms are called \emph{authentication factors}, and typical examples include, \emph{inter alia}, knowledge factors (e.g., passwords, pin codes, etc.), possession factors (e.g., verification code via text message or email), inherent factors (e.g., biometrics --- fingerprints, voice/face recognition, etc.). We model a simple MFA example of user-login for an account at a fictional bank called \emph{Bunched Money}, which abstracts banking apps quite generally. 

To access their Bunched Money account, a user needs two out of the three possible authentication factors: a password $\userpwd$, a one-time passcode $\OTP$, or a hardware fob $\hwfob$. Access is denoted by a security token $\access$. This policy can be expressed by a theory $\Gamma$ that consists of exactly one formula: 
\begin{equation*}
    \Gamma := ((\userpwd \mand \OTP) \lor (\userpwd \mand \hwfob) \lor (\OTP \mand \hwfob)) \mto \access
\end{equation*}
The $\mand$ (as opposed to $\land$) because it is crucial that the two authentication factors (e.g., $\userpwd$ and $\OTP$) are from \emph{separate} sources. The separation is essential, as otherwise these two factors would not enhance the security compared to a single factor --- for example, two passwords are not substantially stronger than one.

Let $\baseB$ be a model of the policy described by $\Gamma$ --- that is, $\baseB$ is defined by the property $\suppBI{\munit}{\baseB}\Gamma$. Let $R$ be an appropriate collection of verification factors --- for example, $R:=\userpwd\mcomma\OTP$. That access is granted under these conditions is expressed by $\suppBI{R}{\baseB}\access$, which indeed obtains.  

\section{Discussion: The Generality of the Approach}
\label{sec:generality}

\begin{figure}[t]
    \hrule 
    \vspace{5mm}
    \centering
    \includegraphics[width=\linewidth]{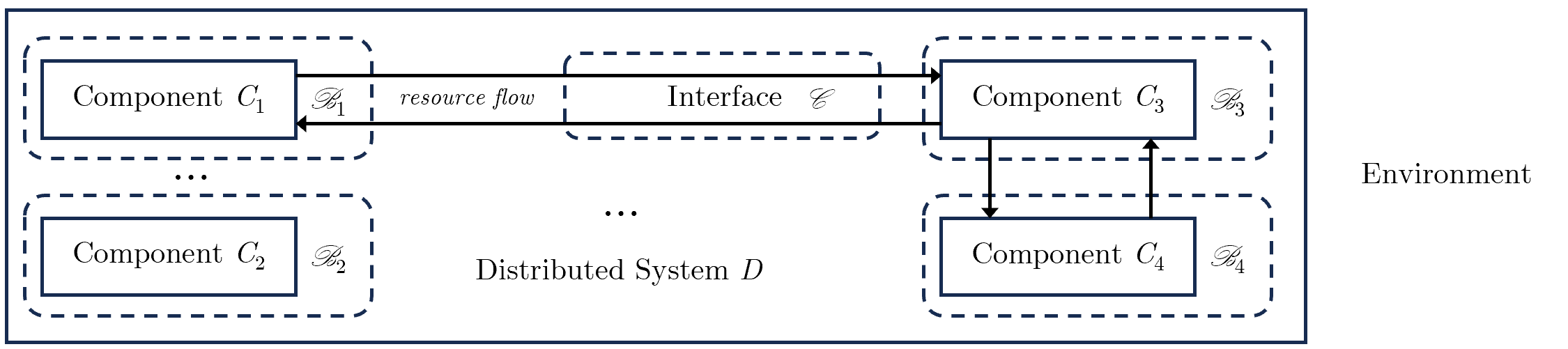} 
    \vspace{1mm}
    \hrule
    \caption{General Distributed System Architecture} 
    \label{fig:generic}
\end{figure}

In Section~\ref{sec:intro}, we discussed an abstract view of (distributed) systems based in the concepts of \emph{location}, \emph{resource}, and \emph{process}: processes manipulate resources that reside at locations. A system is located within an \emph{environment}, itself a system. In general, one can sketch a system with diagrams as in Figure~\ref{fig:generic}; for example, the modelling of airport security in Section~\ref{sec:example} begins with the sketch in
Figure~\ref{fig:airport}.

What we have offered in Section~\ref{sec:res-int} is an interpretation of the B-eS of ILL and BI in terms of systems concepts. We now explain how it applies for modelling systems described by a diagram as in  Figure~\ref{fig:generic}. 

Relative to a diagram such as Figure~\ref{fig:generic}, we can explain quite generally how the inferentialist resource semantics applies to systems modelling.  A \emph{distributed} system $D$ comprises distinct components (or subsystems) $C_1,\ldots,C_n$, each of which has its own policy which may interface with each other. Using the resource interpretation in Section~\ref{sec:res-int}, the policy at the $C_i$ are individually described by formulae $\phi_i$ and the policy of the distributed system $D$ is described by a bunch $\Gamma$ over these descriptions. An inferential \emph{model} of the policy of $D$ is given by a base $\baseB$ supporting $\Gamma$. 

\begin{remark}
    Following the resource semantics of bunches,  one can model each component's policy $\phi_i$ individually by a base $\baseB_i$, and construct a model $\base{D}$ of the overall policy for $D$ by taking the union of those bases together with some rules $\baseC$ governing their interfacing, $\base{D}:=\baseB_1 \cup \ldots \cup \baseB_n \cup \baseC$. In this sense, the resource semantics given by B-eS is \emph{compositional}.
\end{remark}

\begin{remark} The inferential models of policies can be given at various degrees of refinement. In modelling  distributed systems, this corresponds to the fact that each component $C_i$ of $D$ may, if useful, be thought of as a distributed system in itself; that is, one \emph{substitutes} $C_i$ for a more detailed set of systems $C_1^{(i)}$,\ldots,$C_k^{(i)}$, which interface to yield the overall effect of $C_i$. 
The base modelling the policy of $C_1^{(i)}$,\ldots,$C_k^{(i)}$ 
still models the policy of $C_i$. No change is needed.
\end{remark}

This account implicitly assumes a formal model of the compositional structure of a distributed system,  as sketched in Figure~\ref{fig:generic}. There are of course many ways of establishing such a model (e.g., \cite{BarwiseSeligman1997,CMP2012,AP16,caulfield2022eng,GLP2024}). For our current purposes, we consider, at an appropriate level of abstraction, a conceptualization of distributed systems in terms of locations, resources, and processes, as introduced in Section~\ref{sec:intro} \cite{CMP2012,AP16,caulfield2022eng,GLP2024}, as follows (though note the consistency of this perspective with that of \cite{BarwiseSeligman1997}, based on `infomorphisms'): 
\begin{itemize}[label=--]
    \item the distributed system $D$ is based on a directed graph (or a similar topological structure) with certain sub-graphs denoting the components $C_i$;
    \item the vertices of the graph denote locations and the edges 
    give the connectivity between the components;
    \item each component system has some collection of resources that carries some algebraic structure (e.g., as a bunch, or an ordered monoid, and so on) that is coherent with its sub-graph structure; and 
    \item processes, perhaps represented using a resource-process algebra (see, for example, \cite{CMP2012,AP16}) or a primitive notion of behaviour (see, for example, \cite{GLP2024}), describe the manipulation of resources and hence the delivery of services by the system.  
\end{itemize}
Relative to such architecture --- in which the states of a system are 
described by processes executing relative to resources that reside at locations --- logical formulae represent the policy of the system; that is, given a certain distribution of resources across the locations, what subsequent distributions are possible. The details of these readings are given in Section~\ref{sec:res-int}; for example, the formula $\phi \lor \psi$ represents the policy that the system may behave as any state satisfying $\phi$ or $\psi$ ambiguously --- that is, it may move into a state satisfying policy $\chi$ if it is the case that were it in a state  $\phi$ it could move into a state $\chi$ and were it in a state  $\psi$ it could move into a state $\chi$. 

Within Figure~\ref{fig:generic}, we suggest two possible situations in describing the resource flow between components: first, resources may be subject to compliance with policy, as between $C_1$ and $C_3$, and so we also model the \emph{interface} between components; second, resource flow may not be subject to compliance with policy, as between $C_3$ and $C_4$, but must sill be compliant with the policy of the components themselves. These situations can be seen to arise in the example depicted in Figure~\ref{fig:airport}. 


 In the first case,  a passenger may move from ground-side to air-side provided they are compliant with the airport's security policy: in Figure~\ref{fig:airport}, $l_3$ can be seen as such an interface between (models of) ground-side and air-side, consisting of policies that determine whether the passenger after check-in (i.e., component $l_1$) could continue to passport control (i.e., component $l_4$). In the second case, the departure gate implements a check of boarding cards and passports, and perhaps also compliance with cabin-baggage policy, but the details of these are not specified as an interface component of the model.

In general, the permitted manipulations of resources are determined by the system's policies and these policies can be described as logical formulae that are interpreted according to the principles of the resource semantics that we have described in general in Section~\ref{sec:res-int} and illustrated by examples in Section~\ref{sec:example}.

Further work, beyond our present scope, is to develop conceptual and modelling tools for this general set-up more formally, 
perhaps starting from a `minimalistic' approach to systems modelling \cite{GLP2024}.

\section{Conclusion}

The concept of a \emph{distributed system} is foundational in informatics, characterizing the architecture of both physical and abstract infrastructures that are an integral part of modern life. Logic serves as a vital tool for representing, understanding, and reasoning about such systems. One common approach is to give `resource semantics' of logics; that is, interpretations of logical structures and relations in terms of system concepts. Notable examples include the number-of-uses interpretation for linear 
logics and the sharing/separation interpretation for bunched logics. 

Despite the distinct nature of these two resource semantics, this paper presents a unified definition of `resource semantics' that encompasses both. Furthermore, it offers a consistent interpretation of the inferentialist perspective across these logics, specifically through their \emph{base-extension semantics} that uniformly recovers their established resource readings. This underscores the paper's thesis, as articulated in Section~\ref{subsec:thesis}, asserting inferentialism as an intuitive 
and useful framework for logical systems modelling. 

The thesis is illustrated through the modelling of airport security architecture and multi-factor authentication systems, which demonstrate the generic applicability of the approach. Future research, discussed in Section~\ref{sec:generality}, includes formalizing the resource interpretation by establishing a conceptual model of distributed systems and verifying the faithfulness and adequacy of the resource interpretation within that model. From a logical point of view, there is room to explore the base-extension semantics of logic models that precisely capture desired properties of distributed systems; for example, potentially incorporating action and knowledge modalities to refine the understanding of resource movement and policy across distributed systems.

In summary, this paper provides a 
conceptually and technically well-grounded starting point for developing a 
general, systematic, logic-based inferentialist semantics for systems modelling.  



\bibliographystyle{entics}
\bibliography{refs}

\end{document}